Discussion Paper:Communication based on unilateral preference on Twitter: Internet luring in Japan(2018)

# Detection Method for Social Subsets Consisting of Anti-Network Construction for Unilateral Preference Behavior on Directed Temporal Networks

Yasuko Kawahata [†]

Faculty of Sociology, Department of Media Sociology, Rikkyo University, 3-34-1 Nishi-Ikebukuro,Toshima-ku, Tokyo, 171-8501, JAPAN.
ykawahata@rikkyo.ac.jp,kawahata.lab3@damp.tottori-u.ac.jp

**Abstract:** In existing research, as an example of one-sided preference, a conversation structure in which a person who is assumed to be an adult mainly sends one-sided messages to a person who is assumed to be a minor was observed. If subgraphs composed based on such unilateral preferences could be automatically extracted from the network structure, it would be possible to automatically detect communication conducted based on specific motivations from a vast amount of conversation data. In this study, we construct a bottom-up method to detect subgraphs composed of unilateral preferences in Greedy, and discuss the subset of unilateral preferences detected in this simulation.

## 1. Introduction

In existing research, as an example of one-sided preference, a conversation structure in which a person who is assumed to be an adult mainly sends one-sided messages to a person who is assumed to be a minor was observed. If subgraphs composed based on such unilateral preferences could be automatically extracted from the network structure, it would be possible to automatically detect communication conducted based on specific motivations from a vast amount of conversation data. However, nowadays, luring is established using various SNSs such as Telegram, Instagram, and Meta (Facebook), in addition to Twitter (X). In this study, we extract a subgraph with unilateral preferences from a cluster of several thousand nodes with multiple conditions and discuss its contents. Complex networks are constructed based on various principles. In many cases, homophiliy and preferential at-tachment are effective in understanding and predicting the overall structure of complex networks. In addition, recent studies have shown that some mesoscopic structures exist in some regions that cannot necessarily be explained by these principles alone when observing the details of the network. Various unilateral preferences are assumed to exist. In particular, in online social networks, it is easy to search for people with specific attributes and to induce their opinions, so one-sided preferences are likely to lead to actual actions such as sending messages. In networks other than OSNs, for example, it is possible to assume behaviors such as liking or attacking people with certain attributes in a conversational network. If we can detect such unilateral preferences, we may be able to detect conversations based on unilateral selection. Our hypothesis is that some of them involve dangerous communication such as "Luring," and observing such things gives us suggestions for the safe use of online social networks. Furthermore, detecting subsets that are considered to be composed based on unilateral preferences from intricate and complex networks may also lead to the estimation of the meaning of links. In this study, we construct a bottom-up method for detecting subgraphs composed of unilateral preferences in Greedy, and we perform a time-series network simulation on the subset of unilateral preferences detected in this simulation, adding new clusters such as anti-networks and block networks. Hypotheses will be made and discussed.

## 2. Previous Research

Much research on community formation created by unilateral network building has been conducted in the areas of social media analysis, information diffusion, online communities, and influence networks. Research on one-way network building is conducted to understand the social relationships found in situations where interaction is not necessarily two-way, such as social media follow functions and citation networks. Such networks are often analyzed as "directional networks," characterized by the unidirectional nature of the relationships between nodes. Network science methods are used to analyze how unilateral preferences and relationships influence community formation. These studies are also important factors in understanding the use of social media platforms, information diffusion mechanisms, and interactions between individuals within a network. Research on social network dynamics aims to understand patterns of people's interactions and how they evolve over time. Network science approaches are also advancing efforts to analyze these dynamics quantitatively, often using complex systems and statistical physics methods to model social processes, hypotheses, and discussions. The study of social network analysis of the spread of misinformation and disinformation is also essential to understanding the influence of social media, patterns of information transmission, and how quickly these can spread. This chapter section summarizes with respect to previous research.

## 2.1 Network Dynamics

Kossinets, G., and Watts, D. J. (2006), Empirical Analysis of an Evolving Social Network, empirically analyzes the dynamics of social networks as they form over time and provides a deep understanding of the evolution of social structures. Centola, D. (2010), The Spread of Behavior in an Online Social Network Experiment, compares the propagation patterns of behavior in an online network between random and planned networks. and experimentally explore the mechanisms by comparing patterns of propagation of behavior in online networks between random and planned networks. Kleinberg, J. (2000), Networks, Dynamics, and the Small-World Phenomenon, analyzes small-world phenomena in networks and explores how preferences and dynamic properties interact. Romero, D. M., Galuba, W., Asur, S., and Huberman, B. A. (2011), Influence and Passivity in Social Media, analyzes the dynamics among users with respect to influence and passivity in social media to gain a better understanding of information flow and its spread. McPherson, M., Smith-Lovin, L., and Cook, J. M. (2001), Birds of a Feather: Homophily in Social Networks, examined the principle of homophily in social networks and found that individuals tend to associate based on similarities. Ugander, J., Karrer, B., Backstrom, L., and Marlow, C. (2011), The Anatomy of the Facebook Social Graph, analyzes the detailed structure of the Facebook Social Graph, analyzes the detailed structure of the social graph and provides insights into user interactions and community formation within the network." Collective dynamics of 'small-world' networks" Watts, D. J. and Strogatz, S. H. (1998) developed a model for a phenomenon called small-world networks. This work lays the groundwork for a better understanding of small-worldness by focusing on network structures that have a high degree of clustering and short average path lengths." Emergence of Scaling in Random Networks" The work of Barabási, A.-L. and Albert, R. (1999) shed light on the characteristics of scale-free networks generated by preferential connection of links. This concept explains the characteristics found in various networks in the real world. Watts, D. J, Dodds, P. S., and Newman, M. E. J. (2002) studied individual identity and information retrieval methods in social networks. This research revealed how people search for information within social settings; Csárdi, G., Strandburg, K. J., Zalányi, L., Tobochnik, J., and Érdi, P. (2007) revealed the pathways of information transfer in social communication networks. information transfer pathways in social communication networks. "Social Network Dynamics and Participatory Politics," Bennett, W. L., Segerberg, A., and Walker, S. (2014), analyzed the impact of digital media on political participation and the dynamics. "The spread of evidence-poor medicine via flawed social-network analysis," Lyons, R. (2011) warned about how flawed social network analysis can spread evidence-poor medicine. Alatas, V., Banerjee, A., Chandrasekhar, A. G., Hanna, R., and Olken, B. A. (2016) empirically studied the impact of Indonesian networks on information aggregation. "The spread of true and false news online," Vosoughi, S., Roy, D., and Aral, S. (2018) compared patterns of truth and false information spread online and found that misinformation tends to spread more quickly. "Misinformation and Its Correction: Continued Influence and Successful Debiasing," Lewandowsky, S., Ecker, U. K. H., Seifert, C. M., Schwarz, N., and Cook, J. (2012) explored the persistent effects of misinformation and effective methods for correcting it. "Rumor Cascades," Friggeri, A., Adamic, L. A., Eckles, D., and Cheng, J. (2014) analyzed how rumors spread through Facebook as a cascade phenomenon. "Structural diversity in social contagion" by Centola, D. and Macy, M. (2007) examined the impact of structural diversity within networks on social contagion and provided insights into health behaviors and adoption of new technologies. "Analyzing the Digital Traces of Political Manipulation: the 2016 Russian Interference Twitter Campaign," in Linvill, D. L. and Warren, P. L. (2018). Russian intervention strategy in the U.S. presidential election in 2016 through social media data.

## 2.2 Statistical Dynamics of Networks

"The role of social networks in information diffusion" Eytan Bakshy et al. (2012) - This study analyzes how information is distributed and shared on Facebook and in particular how one-sided relationships It examines how one-sided relationships in particular contribute to patterns of information diffusion. Guille et al. (2013) - This study provides a comprehensive investigation of how information spreads within online social networks, with a particular focus on how one-sided relationships within a network can influence this. Weng et al. (2010) - studies methods for identifying communities that share common interests on Twitter, tracking follower behavior to understand community dynamics. "Sampling from Large Graphs" Leskovec and Faloutsos (2006) - developed a method for efficiently sampling from large graphs to explore the structure and dynamic nature of social networks. "Com-munity Structure in Directed Networks" Arenas et al. (2007) - revealed the community structure within directed networks and demonstrated the effectiveness of the new method by applying it to real-world network data. "Statistical mechanics of complex networks" Albert and Barabási (2002) - dealt with the statistical physics of complex networks and in particular evaluated the properties of scale-free networks in terms of directivity. "Robustness and fragility of Boolean models for genetic regulatory networks" Albert and Othmer (2003) - modeled genetic regulatory networks and analyzed how directed networks analysis of how directed networks affect their robustness and fragility. "Directed network modules" M. E. J. Newman (2008) - applied the concept of modularity to directed networks and proposed a new method for detecting modularity in networks. "Clustering in complex directed networks" De Meo et al. (2014) - provided a detailed analysis of the phenomenon of clustering in directed networks and improved our understanding of the local structure of networks. "Analysis of Large-Scale Social and Information Networks" Jure Leskovec (2008) - analyzed large-scale social and information networks, providing new insights into community formation and information diffusion. Réka Albert and Albert-László Barabási (2002), "Statistical mechanics of complex networks":Albert and Barabási applied statistical physics to analyze the growth and scaling laws of networks, including including the laws of network growth and scaling.M. E. J. Newman, S. H. Strogatz, and D. J. Watts (2001), "Random graphs with arbitrary degree distributions and their applications". Newman and colleagues studied random graphs with arbitrary degree distributions and elucidated the statistical dynamics of directed and undirected networks. M. E. J. Newman (2003), "The structure and function of complex networks":. Newman provides a detailed review that unravels the structural features shared by various networks and their functions. Albert-László Barabási and Réka Albert (1999), "Emergence of Scaling in Random Networks":. Barabási and Albert explored how preferential link formation creates scale-free properties in network growth. Duncan J. Watts and Steven H. Strogatz (1998), "Collective dynamics of 'small-world' networks":. Watts and Strogatz quantitatively analyzed the properties of small-world networks and revealed their statistical dynamics. S. Boccaletti, V. Latora, Y. Moreno, M. Chavez, and D.-U. Hwang (2006), "Complex networks: structure and dynamics". Boccaletti et al. provided an extensive review on the structure and dynamics of complex networks and analyzed the statistical dynamic properties of various networks, including scale-free. Kwang-Il Goh, Byungnam Kahng, and Doochul Kim (2001), "Universal behavior of load distribution in scale-free networks". Goh and colleagues studied and provided insight into the universal behavior of load distribution in scale-free networks.

## 2.3 Percolation Networks

Mathematical framework for understanding phenomena such as phase transitions and cluster formation on random graphs. The application of percolation theory in network science is particularly useful for studying network connectivity, resilience, and the spread of contagion. Network resilience is also a concept that measures how resistant a network is to random or deliberate attacks and how quickly it can regain functionality after a failure. Here are some examples of network science research on random graphs based on percolation theory. Understanding how the topology of a network affects its robustness helps to develop strategies for designing more resilient networks. Network connectivity is also an important property that measures how directly or indirectly accessible each node in a network is to each other. Understanding network connectivity provides insight into how networks should be built and maintained and what vulnerabilities exist. Network connectivity has many practical applications, including information diffusion, traffic flow, social interaction, and the structure of the Internet. It provides a foundation for understanding the complex dynamics of systems involving multiple interdependent networks, such as power and communication networks, economic systems, and social and technological infrastructure. The study of interdependent networks is also important for developing strategies to increase the resiliency of these systems. Targeted attacks refer to attacks that intentionally select the most critical nodes or links in a network to cause failure. Network robustness is also a debate about the resistance of the network to such attacks. The goal is to identify structural weaknesses in the network and to develop strategies for more robust network design based on these weaknesses. The study of network robustness is essential in a wide variety of application areas, including information technology, social networks, economic systems, and infrastructure. Percolation processes in interdependent networks are studied to understand how failures in one network affect other networks. The focus in this area is also generally on cascading failures and chain reactions due to interdependence. These studies provide a better understanding of interdependent networks and how they can improve the robustness of infrastructures such as power grids, communication networks, transportation systems, and financial systems. The study of interdependent networks in network science plays an important role in reducing the risks faced by social and technological systems and in designing more resilient systems.

M. E. J. Newman, S. H. Strogatz, and D. J. Watts (2001) This review paper studied the properties and applications of random graphs with arbitrary degree distributions and derived various network properties of these graphs utilizing generating functions M. E. J. Newman ( 2003) This review paper extensively examined structural and functional properties of complex networks characterized by a variety of de-gree distributions; M. Molloy and B. Reed (1995) provided a method for generating random graphs based on specified degree distributions, and their approach is S. Boccaletti, V. Latora, Y. Moreno, M. Chavez, and D.-U. Hwang (2006) provided a comprehensive review of the structure and dynamics of complex networks, including network properties of different degree distributions. Aaron Clauset, Cosma Rohilla Shalizi, and M. E. J. Newman (2009) developed methods for identifying and evaluating power-law distributions found in real data, which are used to analyze networks exhibiting power-law degree distributions. Réka Albert, Hawoong Jeong, and Albert-László Barabási (2000) Research on the robustness and vulnerability of scale-free networks, focusing on their resilience to random failures and the impact of attacks on specific nodes. Réka Albert, Hawoong Jeong, and Albert-László Barabási (2000) A study of network fault tolerance and vulnerability to attacks on hub nodes. Sergey V. Buldyrev, Roni Parshani, Gerald Paul, H. Eugene Stanley, and Shlomo Havlin (2010) Examined cascading fault effects among interdependent networks. Michelle Girvan, Duncan S. Callaway, Mark E. J. Newman, and Steven H. Strogatz (2002) studied how network topology affects its reactivity. Reuven Cohen, Keren Erez, Daniel ben-Avraham, and Shlomo Havlin (2000) analyzed how resistant the structure of the Internet is to random failures. Duncan S. Callaway, M. E. J. Newman, Steven H. Strogatz, and Duncan J. Watts (2000) Investigated the robustness of random graphs using percolation theory.

Callaway, D. S., Newman, M. E. J., Strogatz, S. H., and Watts, D. J. (2000) - This study used percolation theory on random graphs to investigate how a failure from a single point in a network can spill over to the entire It explored how networks can become fragile or, conversely, how they can remain resilient. Cohen, R., Erez, K., ben-Avraham, D., and Havlin, S. (2000) - Analyzing how resistant the Internet is to random disruption using percolation theory to reveal its structural toughness. Buldyrev, S. V., Parshani, R., Paul, G., Stanley, H. E., and Havlin, S. (2010) - In interdependent network systems, how the effects of a partial network collapse can spread up the chain, percolation model to gain a better understanding of large-scale system disruptions. Fortunato, S. (2010) - provides an extensive review of key phenomena occurring within complex networks, specifically integrating knowledge of percolation and phase transitions. Stauffer, D. and Aharony, A. (1992) - A basic reference describing central concepts in percolation theory and their application on random graphs. Watts, D. J. and Strogatz, S. H. (1998) - pioneering work on population dynamics of small-world networks, revealing a network structure that balances high clustering coefficients and short average path lengths. Barabási, A.-L. and Albert, R. (1999) - Innovative work on scale-free networks and their growth mechanisms, proposing preferential connectivity among network nodes. Tangmunarunkit, H., Govindan, R., Jamin, S., Shenker, S., and Willinger, W. (2002) - focused on the robustness of Internet topology and how its structure resists failure. Fagnant, D. J. and Kockelman, K. (2014) - evaluated the interdependency of transportation networks under targeted attacks and their impact on robustness and discussed critical infrastructure vulnerabilities.

## 3. Network Building

In this method, we want to generate and simulate a type of data with the attributes $A$, $B$, $C$, $D$, and $E$. $A$ is a cluster that keeps sending information unilaterally, $B$ is a cluster that keeps receiving information unilaterally, $C$ is a cluster that overlooks the exchange between $A$ and $B$ and spreads the information, $D$ is a cluster that blocks information from $A$, $B$, and $C$, and $E$ is a cluster that tries to stop sending information by issuing an alert to $A$ that keeps sending information unilaterally. $E$ is a cluster that tries to stop sending information by issuing an alert to $A$, who keeps sending information unilaterally. The order of the number of people is $B > D > A > C > E$. In addition, the amount of information sent out is assumed to be $A > E, E > A, C > B > D$.

$A$ sends information unilaterally to the other clusters. This means that a user in cluster $A$ has an edge over a user in another cluster. Cluster $B$ receives information unilaterally, but does not send it. This means that a user in cluster $B$ has an incoming degree but not an outgoing degree. Cluster $C$ observes and disseminates information from $A$ and $B$. This means that a user in cluster $C$ receives information from $A$ and spreads it to other clusters. Cluster $D$ blocks information. This means that cluster $D$ has no edges from other clusters. Cluster $E$ alerts cluster $A$ and tries to get it to stop disseminating information. This means that cluster $E$ has a specific edge to cluster $A$.

The goal is to find the largest connected components of the network and to identify and visualize $k$-cores. $k$-cores are the largest subgraphs generated such that each node in the network has at least $k$ edges, and consider patterns other than one-way communication.

We also subject the behavior of timesteps to conditions on the behavior on temporal networks on a per-$A$-$E$ basis. For each timesteps

$A$: $A$'s behavior is to keep sending messages unilaterally.

$B$: $B$ continues to respond to some of $A$'s information.

$C$: $C$ keeps overhead.

$D$: $D$ is designed to reinforce the block.

$E$: $E$ wants to behave in a way that builds a hostile network, such as sending information unilaterally to $A$.

Of course, $A$ will ignore $E$, but if more than 15 pieces of information come to one agent during timesteps, then $A$ will stop its behavior. The model for building them is described below.

article amsmath amsfonts

## 3.1 Building of a Cluster Network

The initial configuration of the clusters is as follows:

> Cluster $A$: Size $n_A$, information sent $i_{\text{out},A}$, information received $i_{\text{in},A}$
>
> Cluster $B$: Size $n_B$, information sent $i_{\text{out},B}$, information received $i_{\text{in},B}$
>
> Cluster $C$: Size $n_C$, information sent $i_{\text{out},C}$, information received $i_{\text{in},C}$
>
> Cluster $D$: Size $n_D$, information sent $i_{\text{out},D}$, information received $i_{\text{in},D}$
>
> Cluster $E$: Size $n_E$, information sent $i_{\text{out},E}$ (a random variable), information received $i_{\text{in},E}$

The communication rules between clusters are as follows:

Cluster $A \rightarrow$ Cluster $B$ : Probability $p_{A \rightarrow B}$

Cluster $C \rightarrow$ Cluster $A$ : Probability $p_{C \rightarrow A}$,

Cluster $C \rightarrow$ Cluster $B$ : Probability $p_{C \rightarrow B}$

Cluster $E \rightarrow$ Cluster $A$ : Probability $p_{E \rightarrow A}$

## 3.2 Calculation of Asymmetry Index

The provided Python function, `Calculate_Asymmetry`, computes the asymmetry index for clusters within a network. The asymmetry index is a measure of the imbalance between the incoming and outgoing links of a cluster.

## 3.3 Parameters and Variables

> **network**: A dictionary representing the network, where each key is a cluster and its value is a list of links emanating from that cluster.
>
> **clusters**: A list or dictionary of clusters present in the network.
>
> **in_degrees**: A dictionary where each key is a cluster and its value represents the number of incoming links ($\deg^-$).
>
> **out_degrees**: A dictionary where each key is a cluster and its value represents the number of outgoing links ($\deg^+$).
>
> **total_in_degree**: The sum of all incoming degrees in the network ($\sum \deg^-$).
>
> **total_out_degree**: The sum of all outgoing degrees in the network ($\sum \deg^+$).
>
> **asymmetry_A**: The asymmetry index for cluster $A$, calculated as the ratio of the sum of $A$'s in and out degrees to the sum of all degrees in the network.

## 3.4 Asymmetry Index Calculation

The asymmetry index for cluster $A$ is calculated using the following formula:

$$\text{asymmetry\_A} = \frac{\text{in\_degrees}[A] + \text{out\_degrees}[A]}{\text{total\_in\_degree} + \text{total\_out\_degree}}$$

This function returns the asymmetry index for cluster $A$, which quantifies the directional imbalance of links in the network.

The **function to generate the network for simulation** creates a list of links from each cluster based on the information output and the communication rules. This function generates links from the sending cluster to the receiving cluster based on the amount of information sent and the corresponding probability.

The expression for network link generation in mathematical terms is as follows:

$$L_{\text{sender} \rightarrow \text{receiver}} = i_{\text{out,sender}} \times p_{\text{sender} \rightarrow \text{receiver}}$$

where $L_{\text{sender} \rightarrow \text{receiver}}$ represents the number of links from the sending cluster to the receiving cluster, $i_{\text{out,sender}}$ is the amount of information sent by the sending cluster, and $p_{\text{sender} \rightarrow \text{receiver}}$ represents the probability of communication from the sending cluster to the receiving cluster.

## 3.5 Calculation of Mutual Friend-ship Density

The function `calculate_mutual_friend-ship_density` is designed to compute the density of mutual friend-ships between clusters in a network. Mutual friend-ship density is defined as the proportion of actual bidirectional links to the total possible links between clusters.

### Parameters and Variables

> **network**: A dictionary representing the network, where keys are clusters and values are lists of links to other clusters.
>
> **clusters**: A dictionary of clusters where each key is a cluster identifier and its value is a dictionary containing attributes such as the cluster's size.
>
> **link_counts**: A dictionary to count the number of links between pairs of clusters.
>
> **mutual_friend-ship_density**: A dictionary that will store the computed density of mutual friend-ships between each pair of clusters.

### Mutual Friend-ship Density Calculation

The mutual friend-ship density between two clusters, $A$ and $B$, is calculated using the formula:

$$\text{Density}_{A,B} = \frac{\text{Count}_{A,B}}{\text{Size}_A \times \text{Size}_B}$$

where:

Count$_{A,B}$ is the number of bidirectional links between clusters $A$ and $B$.

Size$_A$ and Size$_B$ are the sizes of clusters $A$ and $B$, respectively.

This density value is computed for each pair of clusters in the network, resulting in a measure of the overall interconnectivity and potential for mutual cooperation between clusters.

## 3.6 Link Count Inspection in Cluster Network

The function `inspect_link_counts_corrected` is intended to compute the actual number of links in a network and to calculate the maximum possible number of links between clusters, discounting self-links.

### Parameters and Variables

**network**: A dictionary representing the network where keys are clusters and values are lists of links emanating from these clusters.

**clusters**: A dictionary containing the size of each cluster.

**actual_link_counts**: A dictionary that records the actual number of links for each cluster.

**total_sizes**: The sum of the sizes of all clusters within the network.

**max_possible_links**: A dictionary that calculates the maximum possible number of links for each cluster, excluding self-links.

### Maximum Possible Links Calculation

The maximum number of possible links for a cluster $A$ is calculated using the formula:

$$\text{MaxLinks}_A = \text{Size}_A \times (\text{TotalSizes} - \text{Size}_A)$$

where:

Size$_A$ is the size of cluster $A$.

TotalSizes is the sum of the sizes of all clusters in the network.

The function returns two dictionaries: one for the actual number of links for each cluster and another for the maximum possible number of links for each cluster.

## 3.7 Corrected Network Generation

The `generate_network_corrected` function creates a network of links between clusters, ensuring that the number of links from cluster $E$ to cluster $A$ does not exceed the maximum possible number of links.

### Parameters and Variables

**clusters**: A dictionary of cluster information, where each key represents a cluster, and the associated value is a dictionary containing attributes such as size and information output rate.

**communication_rules**: A dictionary defining the probability of a link from one cluster to another.

**network_links**: A dictionary that will store the lists of links for each cluster.

**total_sizes**: The sum of sizes of all clusters.

**max_possible_links**: The maximum possible number of links from one cluster to all others, excluding self-links.

### Link Generation

Links are generated based on the output rate of information from each cluster, with special rules for links from cluster $E$ to $A$ to ensure they do not exceed the maximum possible:

num_links = min (sender_info_out, max_possible_links)×probability

where:

sender_info_out is the information output rate for the sending cluster.

probability is the probability of a link from the sending to the receiving cluster.

max_possible_links is calculated as:

$$\text{max\_possible\_links} = \text{Size}_E \times (\text{TotalSizes} - \text{Size}_E)$$

The function returns the modified network links dictionary, reflecting the corrected link generation procedure.

## 3.8 Directed Graph for Information Flow in Clusters

This script models the flow of information among five clusters, each with a unique role in the communication network.

### Cluster Definitions and Roles

Cluster $A$: Sends information unilaterally.

Cluster $B$: Receives information unilaterally.

Cluster $C$: Oversees and disseminates information.

Cluster $D$: Blocks information.

Cluster $E$: Issues alerts to Cluster $A$.

## 3.9 Network Generation

A directed graph $G$ is constructed with nodes representing the users and directed edges representing the flow of messages according to predefined rules.

## 3.10 Degree Distributions

The script calculates the in-degree distribution for Cluster $B$ and the out-degree distribution for Cluster $A$, representing the number of incoming and outgoing messages, respectively. Histograms are generated to visualize the distributions of in-degrees for set $T$ (Cluster $B$) and out-degrees for set $R$ (Cluster $A$).

**Case of Distributions of in-degrees and out-degrees for clusters $B$ and $A$**

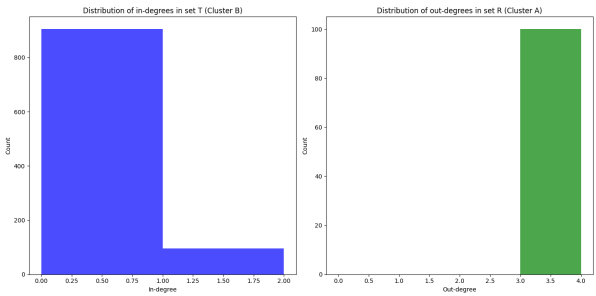

Fig. 1: Histograms showing the distributions of in-degrees and out-degrees for clusters $B$ and $A$, respectively.

**Revised Edge Creation Rules**

The given script generates a directed graph representing a communication network among users in different clusters. The script resets the network, adds nodes for each user in the clusters, and creates edges based on updated rules.

**Network Reset and Node Addition**

The network is cleared and re-initialized as a directed graph $G$. Nodes are added for each user within the clusters $A$, $B$, $C$, $D$, and $E$.

**Edge Creation Rules**

> Cluster $A$ sends messages to clusters $B$, $C$, $D$, and $E$.
>
> Cluster $C$ disseminates information to clusters $B$, $D$, and $E$, assuming it receives information from $A$.
>
> Cluster $E$ sends alerts to cluster $A$.

The edges are added by dividing the number of messages from $A$ and $C$ by the target cluster sizes and randomly selecting nodes within those clusters to create edges. Cluster $E$ adds a random number of edges to $A$, representing alerts.

**Degree Distributions**

For clusters $C$, $D$, and $E$, the script calculates the in-degree and out-degree distributions. The in-degree represents the number of incoming messages, while the out-degree represents the number of messages sent.

**In-degree and out-degree distributions for clusters $C$, $D$, and $E$**

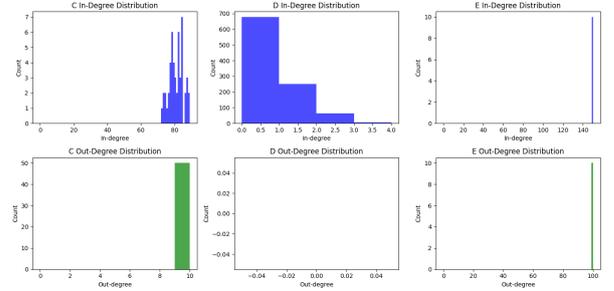

Fig. 2: Histograms showing the in-degree and out-degree distributions for clusters $C$, $D$, and $E$.

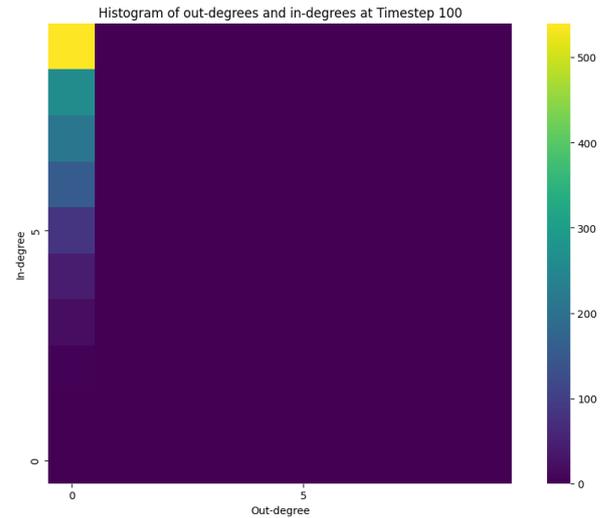

Fig. 3: Histogram of out-degrees and in-degrees at Timestep $t = 100$.

**(1) Considerations on Cluster Behavior**

> **Cluster with risky behavior:** Cluster $E$ is considered to have risky behavior because it is transmitting a large amount of randomly fluctuating information. It may be transmitting a warning or urgent message with a high degree of uncertainty.
>
> **Cluster with unilateral inductive behavior:** Cluster $A$ is considered to be strongly influential because it is unilaterally transmitting a great deal more information than the other clusters. Such behavior may play an important role in shaping opinions.

**Cluster acting defensively:** Cluster *D* receives very little information and its output is similarly limited and can be interpreted as acting defensively. This cluster appears to be blocking information and may upset the balance of information distribution.

### (2) Information Selection Behavior of Passive Cluster *B*

Cluster *B* is passive because it receives a lot of information but transmits very little. To avoid information overload for this cluster, the following strategies should be taken:

a. **Assess importance:** Assess the importance of the information received and filter out irrelevant or unnecessary information.

b. **Aggregate and summarize information:** Aggregate and summarize to effectively process large amounts of information.

c. **Selection of reliable sources:** Focus on information from reliable sources to ensure the accuracy of information.

### (3) Consideration of Social Clusters and Patterns of Opinion Formation

**Central to opinion formation:** Cluster *C* is responsible for monitoring and disseminating information, and this group may play a central role in the opinion formation process.

**Information Barrier:** Cluster *D* may be responsible for blocking the flow of information, possibly creating a barrier to opinion formation.

**Asymmetric information exchange:** Asymmetric information exchange between clusters *A* and *E* may be used to favor some opinions over others or to arouse a sense of caution.

**Effects of information overload:** Cluster *B* receives a large amount of information and can serve as a model for how information overload affects opinion formation. Without proper filtering and summarization, there is a risk of being influenced by misinformation and propaganda.

### (4) Functional Importance of Each Cluster in Opinion Formation

**Leadership:** Cluster *A* may play a leadership role in opinion formation.

**Filtering and gatekeeping:** Cluster *C* may have a gatekeeper role that functions to filter information and ensure that appropriate information is spread throughout society.

**Diversity of opinion:** Cluster *E* could provide different viewpoints at risk and ensure diversity of opinion.

### (5) Characteristics of Inferred Clusters

**Cluster *A*: Media organizations and influencers**

- **Characteristics:** This type is inferred to have a strong influence on opinion formation by disseminating a large amount of information.
- **Possible case:** Mainstream media or well-known social media influencers disseminating information that determines the flow of discussion. These sources will often take a leadership role in shaping public opinion.

**Cluster *B*: The general public**

- **Characteristics:** It is inferred that they are more active in receiving information, but less in disseminating it.
- **Possible case:** consumers and general readers of social media receive information from a wide variety of sources, but do not transmit much themselves. They play the role of receivers in the opinion formation process.

**Cluster *C*: Experts and critics**

- **Characteristics:** Monitoring, selection, and dissemination of information.
- **Possible case:** Experts and critics select important information, interpret and evaluate it, and disseminate it to the public. They are presumed to play a filtering and gatekeeping role and ensure the quality of information.

**Cluster *D*: Private groups or closed communities**

- **Characteristics:** They engage in limited information exchange and are presumed to be defensive against information from outside.
- **Possible cases:** Communities where information circulates in specific online forums or closed groups, and where there is a skeptical or hostile attitude toward outside information sources. These may form echo chambers.

**Cluster *E*: Activist or fringe groups**

- **Characteristics:** Risky dissemination of information to stimulate debate.
- **Potential Cases:** Political and social activists and groups with views outside the mainstream share their own information and interpretations, challenging existing debates. It can be inferred that these are often the spark that creates the debate.

These clusters play different roles in information flow and opinion formation, but with varying degrees of influence in their respective social and cultural contexts. The interaction between these clusters is important in the opinion formation process, especially as the reliability, bias, and accessibility of information can have a significant impact on public opinion. In the context of social debate and policy making, it is important to understand the influence of these clusters and to develop appropriate communication strategies.

# 4. Dynamic Network Simulation and Degree Distribution

This simulation initializes a dynamic network with nodes representing users in various clusters. Edges are added to the network to simulate communication based on certain rules and are updated at each timesteps.

## Network Initialization

A directed graph $G$ is created to represent the network. The initial sizes of the clusters and the message sending behavior are defined as follows:

$$\text{Cluster Sizes:} \{A, B, C, D, E\}$$
$$\text{Random between 100 and 10000}\}$$

Nodes are added to graph $G$ for each user in clusters $A$ through $E$.

## Network Update at Each Timestep

At each timestep, the network is updated:

Cluster $E$ sends alerts to $A$. If a user in $A$ receives 15 or more messages, they stop acting.

Cluster $A$ sends messages to all other clusters unless stopped by alerts from $E$.

## Degree Distribution

Every 100 timesteps, the in-degree and out-degree distributions for each cluster are visualized:

The in-degree represents the number of incoming messages to a node.

The out-degree represents the number of messages sent from a node.

Degree distributions are visualized using histograms for both in-degrees and out-degrees at specified timesteps. The main simulation loop runs for 1000 timesteps, calling the update function at each step.

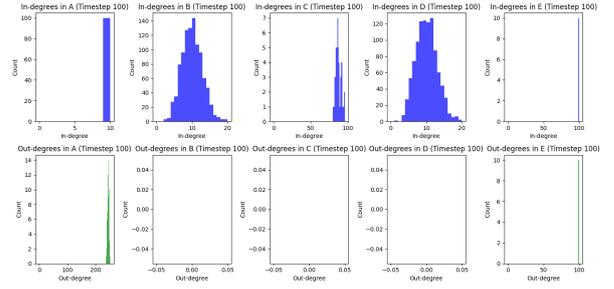

Fig. 4: Histograms showing the distributions of in-degrees and out-degrees for clusters, respectively, $T = 100$

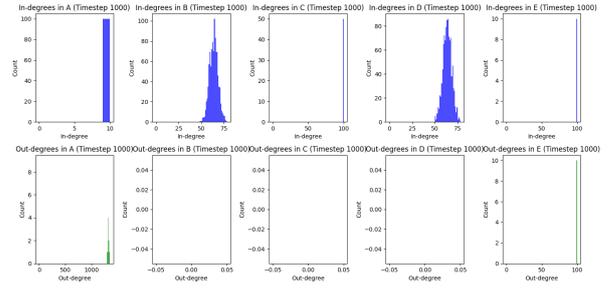

Fig. 5: Histograms showing the distributions of in-degrees and out-degrees for clusters, respectively, $T = 1000$

## Case of Distributions of in-degrees and out-degrees for clusters

### (1) Considerations

At time steps $t = 100$, 900, and 1000, clusters $A$ and $E$ have very high incoming orders but very low outgoing orders. This means that these clusters receive information extensively but transmit very little ($A$ is for transmission only and $E$ is for warning only).

The distribution of incoming orders for clusters $B$ and $D$ varies with time, with many nodes having several incoming orders. This indicates that a large number of nodes are receiving information, but because $D$ blocks that information, its distribution is flatter than $B$'s.

Cluster $C$ is responsible for monitoring and spreading the information, so a certain number of nodes have a high incoming degree, suggesting that it is an information aggregation point. Since the distribution of outgoing orders is almost zero, it is likely that these nodes play the role of relaying information.

The slight increase in the outgoing order counts of clusters $A$ and $E$ at time step 1000 suggests that the nodes in these clusters have become active over time.

The change in the shape of the incoming order distribution for cluster $B$ at each time step indicates a change in the pattern of information reception within this cluster, perhaps reflecting the diffusion of information with time.

While this data provides insight into the dynamic behavior of the network and the flow of information, details regarding the specific network topology and interactions among the nodes are unknown. Further analysis would require information on the specific structure of these networks and the relationships among the nodes.

**(2) Possible scenarios**

**Political propaganda:**

During an election period, a particular politician or political party manipulates information to emphasize its own platform, spreading news to arouse suspicion and distrust.

Important information that could influence the election could spread quickly through social networks and have a significant impact on the outcome of the election.

**Manipulation of stock prices in the market:**

Cases in which false information is disseminated that could affect stock prices, resulting in unfair manipulation of stock prices.

The dissemination of such information is highly sensitive because the information shared among investors has a significant impact on the market and the accuracy of the information is directly related to asset values.

**Diffusion of Social Movements:**

This is a case of using social media to spread the message of a social movement. For example, environmental protection activities or human rights movements.

The accuracy and attractiveness of the information can affect the speed and scope of the spread, leading to movements that promote social change.

**Health and Medical Information Sharing:**

Cases in which information about new health risks or diseases is shared among health professionals and the general public.

Accuracy and immediacy are especially important, as this information directly influences people's health behavior.

**Information Exchange on Education:**

This is when information on educational reforms and new educational technologies is shared among educators and parents.

Information on the quality of education has long-term implications for future generations, so its dissemination must be done with caution.

These scenarios illustrate the value of information and how its dissemination can have social, economic, and political consequences. However, the discussion here is only hypothetical, and in actual cases, diverse factors are involved in complex ways.

## 5. Discussion

The script performs an analysis on a directed network graph $G$ by identifying its largest weakly connected component and calculating its k-core decomposition. Network is then created, highlighting the k-core structures.

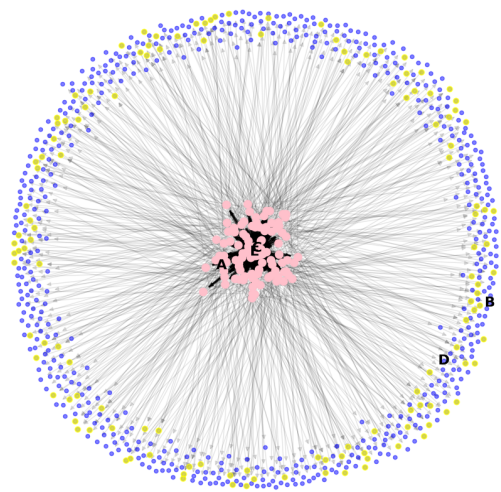

Fig. 6: Case 1:The Largest Connected Component with k-core (k=2 in yellow, k=6 in pink)

**(1) Cluster Behavior Considerations**

**Cluster with risky behavior:** Class $E$ is defined as an alert cluster, which may send random and large amounts of messages to Class $A$ (100 to 10,000). This randomness and variation in message volume can be viewed as risky behavior. In addition, the fact that they do not appear in the high $k$-core region (pink) suggests that they are likely to be located outside the mainstream network.

**Cluster with one-sided inductive behavior:** Class $A$ is the cluster that sends information unilaterally and transmits the largest amount of messages. Since their $k$-core is located in the near-center region (pink), they have a strong influence in the network and control information flow to the other clusters. This can be considered to be the case.

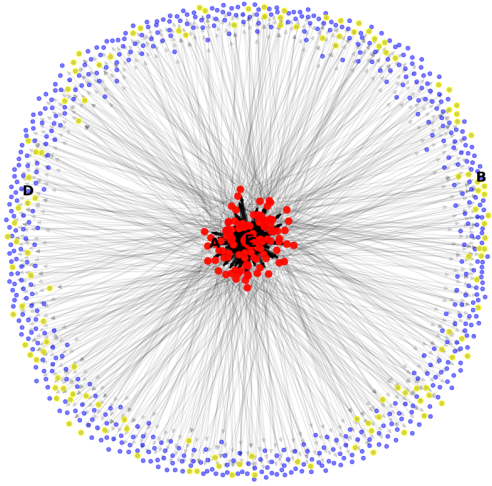

Fig. 7: Case 2:The Largest Connected Component with k-core (k=2 in yellow, k=6 in red)

**Cluster with defensive behavior:** Class *D* is an information blocking cluster, which limits its own information exchange and is defensive against external information. Although its position cannot be clearly confirmed from the graph, it is expected to have few ties with other clusters.

**(2) Information Selection Behavior to be Taken by Passive Cluster *B***

Class *B* is a passive cluster that receives information but does not transmit much. The information selection behavior they should take is as follows:

(1) **Diversification of information sources:** They should take in information not only from Class *A* but also from Classes *C* and *E* to ensure the diversity of information sources.

(2) **Critical evaluation of information:** Critically analyze and fact-check the information provided, rather than relying on it.

(3) **Avoid echo chambers:** Do not stick to a particular source of information or opinion, but listen to different perspectives and opinions.

**(3) Consideration of Social Clusters and Patterns of Opinion Formation Assumed from the Graph Results**

The following patterns of social opinion formation can be read from the graph:

**Centralized opinion formation:** It can be seen that Class *A* plays a central role in disseminating a large amount of information and exerting a significant influence on opinion formation.

**Role of experts and gatekeepers:** Class *C* is responsible for sorting and disseminating information and may contribute to ensuring the quality of information.

**Information blocking and echo chambers:** Class *D* is expected to circulate information within its own information sphere and be closed to outside information. This may reduce diversity of opinion and create echo chambers.

**Information monitoring and alerting:** The presence of Class *E* indicates vigilance toward certain information sources and may contribute to correcting biased information flow. However, its randomness may also cause excessive alarm and anxiety.

These patterns are phenomena that can also be found in real-world media ecosystems and social networks, and a balance between these clusters is important for maintaining a healthy public space.

## Case of Hard Distributions of in-degrees and out-degrees for clusters

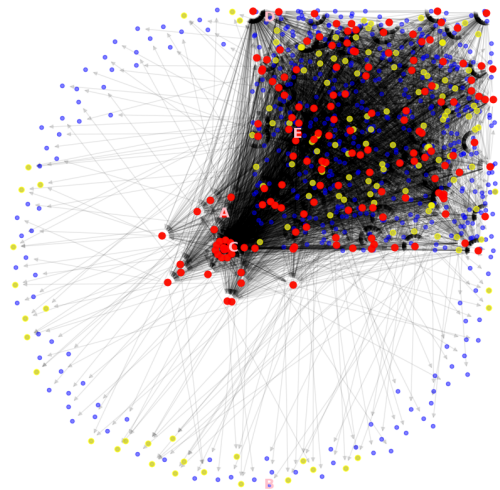

Fig. 8: Case 3:The Largest Connected Component with k-core (k=2 in yellow, k=6 in red)

**(1) Considerations on Cluster Behavior**

**Cluster with risky behavior:** Class *A* is considered a "one-way information dissemination cluster" and sends a large volume of messages. This cluster appears to prioritize quantity over quality and accuracy of information, and is at risk of spreading misinformation and propaganda.

**Cluster with one-sided, inductive behavior:** Class *C* is considered an "information monitoring and dissemination cluster," but it deals with large amounts of information in a relatively small size. This may act as a gatekeeper for opinion formation and information distribution.

**Cluster acting defensively:** Class *D* is considered an "information blocking cluster" and despite its large size, it sends out few messages. This cluster appears to be preventing an overflow of information or blocking certain information. Class *E* is also a small elite "warning cluster against *A*" and seems to play the role of sounding the alarm against the risk of information emitted by *A*.

### (2) Information Selection Behavior of Passive Cluster *B*

Class *B* is considered a "one-way information receiving cluster." This cluster may not play an active role in selecting the quality of information and is therefore susceptible to information overload and misinformation. Therefore, more active information literacy behaviors are required, such as scrutinizing information, checking sources, and gathering information from diverse perspectives.

### (3) Consideration of Social Clusters and Patterns of Opinion Formation

The social clusters assumed from this network diagram seem to function as information senders, diffusers, receivers, defenders, and warners. The pattern of opinion formation is that there is a strong source of information (Class *A*), which spreads through a specific monitoring and diffusion cluster (Class *C*). On the other hand, receiving clusters (Class *B*) are more likely to accept information without selection, and a certain balance is maintained by defensive clusters (Class *D*) and risk warning clusters (Class *E*) to deal with an overflow of information. This structure may contain risks such as concentration of power based on information, formation of echo chambers, and dissemination of misinformation. In addition, we can see how the various clusters interact and influence each other, contributing to the formation of the opinions of society as a whole.

### (5) Characteristics of Inferred Clusters

**Class *A* (one-way information dissemination cluster):** This cluster may correspond to mainstream media or large social media influencers. They can unilaterally disseminate large amounts of information to the masses and can have a significant impact on public debate and opinion formation. However, they carry the risk of spreading misinformation or biased speech, as accuracy and balance of information cannot always be guaranteed.

**Class *B* (one-way information receiving cluster):** Can be viewed as the average consumer or passive recipient of information. This cluster is exposed to large amounts of information, but may be less likely to actively filter or critically analyze it. These recipients are more likely to have their opinions formed by pop culture, advertising, or political propaganda.

**Class *C* (Information Monitoring and Dissemination Cluster):** Experts, critics, and fact-checkers would fall into this category. These actors are expected to monitor and interpret information and provide quality information for social dialogue. However, their influence depends on their reliability and reach, so the information they disseminate must also be carefully verified.

**Class *D* (Information Blocking Cluster):** This cluster may correspond to cases where a government censorship agency or certain companies seek to control information exclusively. The purpose is to suppress certain speech or prevent the spread of certain information by restricting the flow of information.

**Class *E* (Warning Cluster against *A*):** may include activists, independent media, or whistleblowers. They serve to provide alternatives to mainstream sources or to issue warnings on specific issues, and although small in size, their influence can be significant in some cases.

These clusters represent the flow of speech in contemporary society, and each cluster differs in the way information is produced, distributed, and received. Many social issues arise from the interaction of these clusters, including freedom of speech, equality of access to information, media diversity, and protection of privacy. In real societies, these clusters are intricately intertwined, shaping the modes of opinion formation in democratic public spaces.

## Largest Weakly Connected Component

The largest weakly connected component (WCC) of the network is identified. A WCC is a subset of nodes where there is a path in either direction between every pair of nodes. Mathematically, it is defined as:

$$\text{Largest WCC} = \max_{\text{all WCCs in } G} |\text{WCC}|$$

The subgraph induced by this component is denoted by $G_{\text{WCC}}$.

## k-core Decomposition

The k-core of a graph is a maximal subgraph in which each node has at least *k* connections within the subgraph. The k-core decomposition is calculated for two different values of $k$ (abstractly represented by $k_1$ and $k_2$), which are defined as:

$k\text{-core}_{k_1}$ = maximal subgraph with degrees $\geq k_1$

$k\text{-core}_{k_2}$ = maximal subgraph with degrees $\geq k_2$

## Node Position Calculation and Visualization

Node positions are calculated using a force-directed layout algorithm (e.g., spring layout) applied to a subsample of the largest WCC to reduce computational cost. Representative nodes from each cluster are identified for labeling purposes.

The network is visualized by plotting:

Nodes and edges of the largest WCC.

Nodes of the $k_1$-core subgraph highlighted in one color.

Nodes of the $k_2$-core subgraph highlighted in another color.

Labels are assigned to representative nodes for each cluster to indicate the cluster to which they belong.

# 6. Conclusion

## 6.1 Preference Behavior on Directed Temporal Network

In the preliminary research for this thesis, we will summarize the final considerations regarding time series analysis that I would like to work on in the future. Sets up a directed network to simulate the propagation of messages through various clusters, each with distinct roles. It also provides a mechanism for visualizing the network's structure at specified intervals.

## Network Initialization

A directed graph $G$ is initialized with a set of clusters, each with a predetermined number of nodes (denoted by $n_A, n_B, n_C, n_D, n_E$) corresponding to different roles in information dissemination.

## Message Propagation

Each cluster is associated with a message generation function. For instance, cluster A generates a large number of messages $m_A$, while cluster E generates a variable number of messages $m_E(t)$, which is a random value at each timestep $t$.

## Network Update

At each timestep, nodes from clusters A and E generate edges directed towards nodes in other clusters, based on their message functions. Cluster C observes and spreads information to clusters D and E, while cluster D acts as an information blockade.

## k-core Analysis

The k-core of a graph is defined as a maximal subgraph in which each node has at least $k$ edges. The script calculates k-cores for multiple values of $k$ within the largest connected component of the graph. For a given $k$, the k-core is denoted by $\text{core}_k(G)$.

Every $X$ timesteps, the largest connected component of the network and its k-cores are visualized. Nodes in the k-cores are highlighted with distinct colors to illustrate the varying levels of connectivity.

## Mathematical Notation

Let $V$ be the set of nodes and $E$ be the set of edges in graph $G$. The update rules can be mathematically described as follows:

For each node $v \in V$ belonging to cluster A, if the out-degree of $v$, $\deg^+(v)$, is less than 15, edges are generated to nodes in clusters B, C, D, and E according to the message count $m_A$.

For each node $v \in V$ belonging to cluster E, edges are generated to randomly selected nodes in cluster A based on the message function $m_E(t)$.

Nodes in cluster C act based on a similar stochastic process, directing edges towards clusters D and E.

## Degree Distributions

At specified timesteps, the in-degree and out-degree distributions of the network are visualized, providing insight into the information flow dynamics.

### (1) Consideration of Cluster Behavior

Cluster $A$ is a one-way information sending cluster, sending a large amount of information (4000) but receiving little. This is considered risky behavior and may cause information overload.

Cluster $B$ is on the receiving end, sending very little information and is easily guided and influenced.

Cluster $C$ monitors and disseminates information and may help filter and verify information.

Cluster $D$ is responsible for blocking information and can be said to act defensively. This prevents excessive diffusion of information.

Cluster $E$ has the role of issuing alerts to Cluster $A$ and can sound the alarm for information in Cluster $A$.

### (2) Information Selection Behavior of Cluster $B$

Cluster $B$ needs to select and consume information. It receives a large amount of information, but how it handles it is important. It is wise to focus on information from Cluster $C$

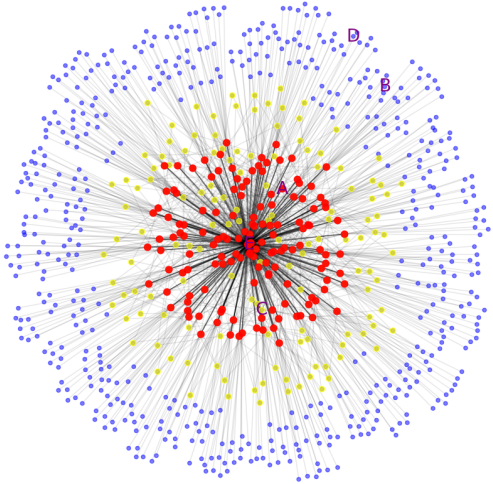

Fig. 9: Case 1, $T = 100$:Largest Connected Component with k-core (k=2, 6, 23) at Timestep

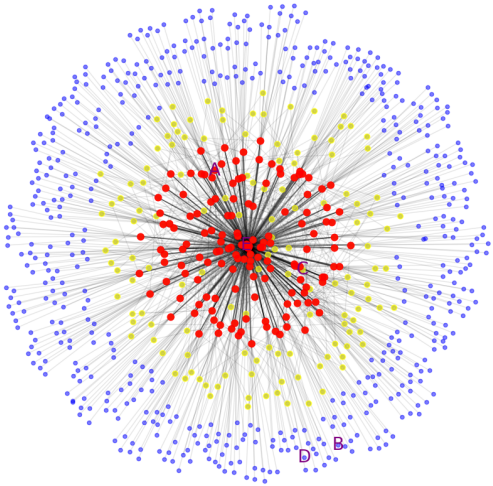

Fig. 10: Case 2, $T = 500$:Largest Connected Component with k-core (k=2, 6, 23) at Timestep

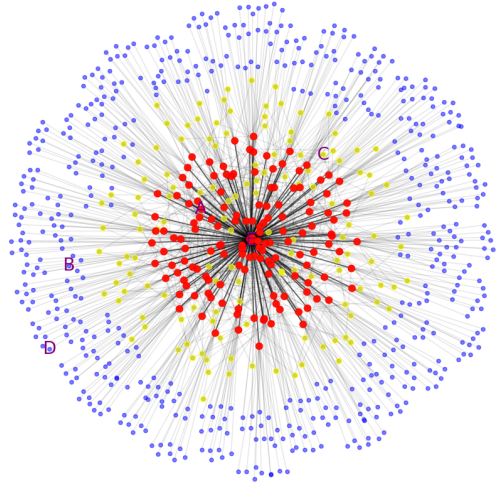

Fig. 11: Case 3, $T = 1000$:Largest Connected Component with k-core (k=2, 6, 23) at Timestep

and use filtering by Cluster *D* to avoid excessive information from Cluster *A*.

### (3) Consideration of Social Clusters and Patterns of Opinion Formation

In society, there is a one-way pattern in which some groups transmit a large amount of information and many other groups receive it. Opinion formation can be influenced by information monitors (Cluster *C*) and blockers (Cluster *D*). Sound opinion formation is aided by monitors improving the quality of information and blockers blocking unnecessary information. A small group of alarmed individuals (Cluster *E*) may play a balancing role in opinion formation.

### (4) Trend of Clusters by Time Step

Cluster activity appears to expand from time step 100 to 1000. This indicates that the information flow is increasing with time. Cluster *A* has been sending large amounts of information from the beginning, and its impact may be expanding over time. In real social systems and information networks, these patterns can provide important insights and a better understanding of how information diffuses and opinions are formed.

### (5) Characteristics of Inferred Clusters

Mass media and large social media platforms (corresponding to cluster *A*): These clusters are characterized by sending out a lot of information in one direction. This applies to certain news channels and social media accounts that produce large amounts of content and distribute it to a wide audience. Over time, the influence of such media may increase and reach even more people.

General audience (corresponding to Cluster *B*): Most people fall into this category; they receive information daily from numerous sources, but disseminate relatively little information. Individuals tend to get trapped in a filter bubble over time and tend to prefer certain types of information.

Experts and Critics (equivalent to Cluster *C*): Experts and critics function as watchdogs of information, evaluating the accuracy and reliability of content and diffusing important information. Over time, experts and critics may become increasingly important sources of information and, depending on their reliability, may increase their influence.

Gatekeepers (equivalent to cluster *D*): Gatekeepers are institutions or persons who control the flow of information, and they curb the spread of information through publication regulation and content moderation. Over time, the influence of these clusters fluctuates as regulations are tightened or relaxed.

Activists and alarmists (corresponding to Cluster *E*): A small group or individual calls society's attention to an issue through raising an issue or issuing a warning. Over time, as society becomes more aware of the issues, the voices of these groups are expected to become stronger.

Changes in information flow and influence among these clusters can be observed in a variety of social contexts, including political campaigns, civic movements, commercial advertising, and educational programs. The behavior and interactions of these clusters will change over time due to various changes in society, such as policy changes, technological advances, and shifting cultural values. Social researchers will monitor these dynamics and hopefully provide insights into maintaining a healthy information ecosystem in society.

# Aknowlegement

This research is supported by Grant-in-Aid for Scientific Research Project FY 2019-2021, Research Project/Area No. 19K04881, "Construction of a new theory of opinion dynamics that can describe the real picture of society by introducing trust and distrust". It is with great regret that we regret to inform you that the leader of this research project, Prof. Akira Ishii, passed away suddenly in the last term of the project. Prof. Ishii was about to retire from Tottori University, where he was affiliated with at the time. However, he had just presented a new basis in international social physics, complex systems science, and opinion dynamics, and his activities after his retirement were highly anticipated. It is with great regret that we inform you that we have to leave the laboratory. We would like to express our sincere gratitude to all the professors who gave me tremendous support and advice when We encountered major difficulties in the management of the laboratory at that time.

First, Prof. Isamu Okada of Soka University provided valuable comments and suggestions on the formulation of the three-party opinion model in the model of Dr. Nozomi Okano's (FY2022) doctoral dissertation. Prof.Okada also gave us specific suggestions and instructions on the mean-field approximation formula for the three-party opinion model, Prof.Okada's views on the model formula for the social connection rate in consensus building, and his analytical method. We would also like to express our sincere gratitude for your valuable comments on the simulation of time convergence and divergence in the initial conditions of the above model equation, as well as for your many words of encouragement and emotional support to our laboratory.

We would also like to thank Prof.Masaru Furukawa of Tottori University, who coordinated the late Prof.Akira Ishii's laboratory until FY2022, and gave us many valuable comments as an expert in magnetized plasma and positron research.

In particular, we would like to thank Prof.Hidehiro Matsumoto of Media Science Institute, Digital Hollywood University. Prof.Hidehiro Matsumoto is Co-author of this paper, for managing the laboratory and guiding us in the absence of the main researcher, and for his guidance on the elements of the final research that were excessive or insufficient with Prof.Masaru Furukawa.

And in particular, Prof.Masaru Furukawa of Tottori University, who is an expert in theoretical and simulation research on physics and mathematics of continuum with a focus on magnetized plasma, gave us valuable opinions from a new perspective.

His research topics include irregular and perturbed magnetic fields, MHD wave motion and stability in non-uniform plasmas including shear flow, the boundary layer problem in magnetized plasmas, and pseudo-annealing of MHD equilibria with magnetic islands.

We received many comments on our research from new perspectives and suggestions for future research. We believe that Prof.Furukawa's guidance provided us with future challenges and perspectives for this research, which stopped halfway through. We would like to express sincere gratitude to him.

We would like to express my sincere gratitude to M Data Corporation, Prof.Koki Uchiyama of Hotlink Corporation, Prof.Narihiko Yoshida, President of Hit Contents Research Institute, Professor of Digital Hollywood University Graduate School, Hidehiko Oguchi of Perspective Media, Inc. for his valuable views from a political science perspective. And

Kosuke Kurokawa of M Data Corporation for his support and comments on our research environment over a long period of time. We would like to express our gratitude to Hidehiko Oguchi of Perspective Media, Inc. for his valuable views from the perspective of political science, as well as for his hints and suggestions on how to build opinion dynamics.

We are also grateful to Prof.Masaru Nishikawa of Tsuda University for his expert opinion on the definition of conditions in international electoral simulations.

We would also like to thank all the Professors of the Faculty of Engineering, Tottori University. And Prof.Takayuki Mizuno of the National Institute of Informatics, Prof.Fujio Toriumi of the University of Tokyo, Prof.Kazutoshi Sasahara of the Tokyo Institute of Technology, Prof.Makoto Mizuno of Meiji University, Prof.Kaoru Endo of Gakushuin University, and Prof.Yuki Yasuda of Kansai University for taking over and supporting the Society for Computational Social Sciences, which the late Prof.Akira Ishii organized, and for their many concerns for the laboratory's operation. We would also like to thank Prof.Takuju Zen of Kochi University of Technology and Prof.Serge Galam of the Institut d'Etudes Politiques de Paris for inviting me to write this paper and the professors provided many suggestions regarding this long-term our research projects.

We also hope to contribute to their further activities and the development of this field. In addition, we would like to express our sincere gratitude to Prof.Sasaki Research Teams for his heartfelt understanding, support, and advice on the content of our research, and for continuing our discussions at a time when the very survival of the research project itself is in jeopardy due to the sudden death of the project leader.

We would also like to express our sincere gratitude to the bereaved Family of Prof.Akira Ishii, who passed away unexpectedly, for their support and comments leading up to the writing of this report. We would like to close this paper with my best wishes for the repose of the soul of Prof.Akira Ishii, the contribution of his research results to society, the development of ongoing basic research and the connection of research results, and the understanding of this research project.# References

[Asatani et al.(2018)] Asatani, K., Kawahata, Y., Toriumi, F., & Sakata, I. (2018). Communication based on unilateral preference on twitter: Internet luring in japan. In *Social Informatics: 10th International Conference, SocInfo 2018, St. Petersburg, Russia, September 25-28, 2018, Proceedings, Part I 10* (pp. 54–66). Springer, **2018**.

[Callaway, D. S., Newman, M. E. J., Strogatz, S. H., Watts, D. J.] Callaway, D. S., Newman, M. E. J., Strogatz, S. H., Watts, D. J.. Network robustness and fragility: Percolation on random graphs. **2000**.

[Cohen, R., Erez, K., ben-Avraham, D., Havlin, S.] Cohen, R., Erez, K., ben-Avraham, D., Havlin, S.. Resilience of the internet to random breakdowns. **2000**.

[Buldyrev, S. V., Parshani, R., Paul, G., Stanley, H. E., Havlin, S.] Buldyrev, S. V., Parshani, R., Paul, G., Stanley, H. E., Havlin, S.. Catastrophic cascade of failures in interdependent networks. **2010**.

[Fortunato, S.] Fortunato, S.. Critical phenomena in complex networks. **2010**.

[Stauffer, D., Aharony, A.] Stauffer, D., Aharony, A.. Percolation theory. **1992**.

[Watts, D. J., Strogatz, S. H.] Watts, D. J., Strogatz, S. H.. Collective dynamics of 'small-world' networks. **1998**.

[Barabási, A.-L., Albert, R.] Barabási, A.-L., Albert, R.. Emergence of Scaling in Random Networks. **1999**.

[Tangmunarunkit, H., Govindan, R. et al] Tangmunarunkit, H, Govindan, R., Jamin, S, Shenker, S, Willinger, W. Robustness of the Internet at the Topology Level. **2002**.

[Fagnant, D. J., Kockelman, K.] Fagnant, D. J., Kockelman, K.. The robustness of interdependent transportation networks under targeted attack. **2014**.

[Albert, R., Barabási, A.-L.] Albert, R., Barabási, A.-L.. Statistical mechanics of complex networks. **2002**.

[Newman, M. E. J., Strogatz, S. H., Watts, D. J.] Newman, M. E. J., Strogatz, S. H., Watts, D. J.. Random graphs with arbitrary degree distributions and their applications. **2001**.

[Newman, M. E. J.] Newman, M. E. J.. The structure and function of complex networks. **2003**.

[Barabási, A.-L., Albert, R.(1999)] Barabási, A.-L., Albert, R.. Emergence of Scaling in Random Networks. **1999**.

[Watts, D. J., Strogatz, S. H.(1998)] Watts, D. J., Strogatz, S. H.. Collective dynamics of 'small-world' networks. (1998).

[Boccaletti, S., Latora, V., Moreno, Y(2006)] Boccaletti, S., Latora, V., Moreno, Y., Chavez, M., Hwang, D.-U.. Complex networks: Structure and dynamics. **2006**.

[Goh, K.-I., Kahng, B., Kim, D.(2001)] Goh, K.-I., Kahng, B., Kim, D.. Universal behavior of load distribution in scale-free networks. **2001**.

[Bakshy et al.(2012)] Bakshy *et al.*. The role of social networks in information diffusion. **2012**.

[Guille et al.(2013)] Guille *et al.*. Information diffusion in online social networks: A survey. **2013**.

[Weng et al.(2010)] Weng *et al.*. Following the follower: Detecting communities with common interests on Twitter. **2010**.

[Leskovec and Faloutsos(2006)] Leskovec and Faloutsos. Sampling from large graphs. **2006**.

[Arenas et al.(2007)] Arenas *et al.*. Community Structure in Directed Networks. **2007**.

[Albert and Barabási(2002)] Albert and Barabási. Statistical mechanics of complex networks. **2002**.

[Albert and Othmer(2003)] Albert and Othmer. Robustness and fragility of Boolean models for genetic regulatory networks. **2003**.

[M. E. J. Newman(2008)] M. E. J. Newman. Directed network modules. **2008**.

[De Meo et al.(2014)] De Meo *et al.*. Clustering in complex directed networks. **2014**.